\documentstyle[12pt]{article}
\begin{document}

\def\ds{\displaystyle}
\def\beq{\begin{equation}}
\def\eeq{\end{equation}}
\def\bea{\begin{eqnarray}}
\def\eea{\end{eqnarray}}
\def\beeq{\begin{eqnarray}}
\def\eeeq{\end{eqnarray}}
\def\ve{\vert}
\def\vel{\left|}
\def\bpll{B\rar\pi \ell^+ \ell^-}
\def\ver{\right|}
\def\nnb{\nonumber}
\def\ga{\left(}
\def\dr{\right)}
\def\aga{\left\{}
\def\adr{\right\}}
\def\lla{\left<}
\def\rra{\right>}
\def\rar{\rightarrow}
\def\nnb{\nonumber}
\def\la{\langle}
\def\ra{\rangle}
\def\ba{\begin{array}}
\def\ea{\end{array}}
\def\tr{\mbox{Tr}}
\def\ssp{{\Sigma^{*+}}}
\def\sso{{\Sigma^{*0}}}
\def\ssm{{\Sigma^{*-}}}
\def\xis0{{\Xi^{*0}}}
\def\xism{{\Xi^{*-}}}
\def\qs{\la \bar s s \ra}
\def\qu{\la \bar u u \ra}
\def\qd{\la \bar d d \ra}
\def\qq{\la \bar q q \ra}
\def\gGgG{\la g^2 G^2 \ra}
\def\q{\gamma_5 \not\!q}
\def\x{\gamma_5 \not\!x}
\def\g5{\gamma_5}
\def\sb{S_Q^{cf}}
\def\sd{S_d^{be}}
\def\su{S_u^{ad}}
\def\ss{S_s^{??}}
\def\ll{\Lambda}
\def\lb{\Lambda_b}
\def\sbp{{S}_Q^{'cf}}
\def\sdp{{S}_d^{'be}}
\def\sup{{S}_u^{'ad}}
\def\ssp{{S}_s^{'??}}
\def\sig{\sigma_{\mu \nu} \gamma_5 p^\mu q^\nu}
\def\fo{f_0(\frac{s_0}{M^2})}
\def\ffi{f_1(\frac{s_0}{M^2})}
\def\fii{f_2(\frac{s_0}{M^2})}
\def\O{{\cal O}}
\def\sl{{\Sigma^0 \Lambda}}
\def\es{\!\!\! &=& \!\!\!}
\def\ar{&+& \!\!\!}
\def\ek{&-& \!\!\!}
\def\cp{&\times& \!\!\!}
\def\se{\!\!\! &\simeq& \!\!\!}
\def\hml{\hat{m}_{\ell}}
\def\rr{\hat{r}_{\Lambda}}
\def\ss{\hat{s}}


\renewcommand{\textfraction}{0.2}    
\renewcommand{\topfraction}{0.8}

\renewcommand{\bottomfraction}{0.4}
\renewcommand{\floatpagefraction}{0.8}
\newcommand\mysection{\setcounter{equation}{0}\section}

\def\baeq{\begin{appeq}}     \def\eaeq{\end{appeq}}
\def\baeeq{\begin{appeeq}}   \def\eaeeq{\end{appeeq}}
\newenvironment{appeq}{\beq}{\eeq}
\newenvironment{appeeq}{\beeq}{\eeeq}
\def\bAPP#1#2{
 \markright{APPENDIX #1}
 \addcontentsline{toc}{section}{Appendix #1: #2}
 \medskip
 \medskip
 \begin{center}      {\bf\LARGE Appendix #1 :}{\quad\Large\bf #2}
\end{center}
 \renewcommand{\thesection}{#1.\arabic{section}}
\setcounter{equation}{0}
        \renewcommand{\thehran}{#1.\arabic{hran}}
\renewenvironment{appeq}
  {  \renewcommand{\theequation}{#1.\arabic{equation}}
     \beq
  }{\eeq}
\renewenvironment{appeeq}
  {  \renewcommand{\theequation}{#1.\arabic{equation}}
     \beeq
  }{\eeeq}
\nopagebreak \noindent}

\def\eAPP{\renewcommand{\thehran}{\thesection.\arabic{hran}}}

\renewcommand{\theequation}{\arabic{equation}}
\newcounter{hran}
\renewcommand{\thehran}{\thesection.\arabic{hran}}

\def\bmini{\setcounter{hran}{\value{equation}}
\refstepcounter{hran}\setcounter{equation}{0}
\renewcommand{\theequation}{\thehran\alph{equation}}\begin{eqnarray}}
\def\bminiG#1{\setcounter{hran}{\value{equation}}
\refstepcounter{hran}\setcounter{equation}{-1}
\renewcommand{\theequation}{\thehran\alph{equation}}
\refstepcounter{equation}\label{#1}\begin{eqnarray}}


\newskip\humongous \humongous=0pt plus 1000pt minus 1000pt
\def\caja{\mathsurround=0pt}

\title{\bf  Black hole thermodynamics and modified GUP consistent with doubly special relativity}
\author{K. Zeynali$^1$\thanks{Email: k.zeinali@arums.ac.ir}
\hspace{2mm},  \hspace{2mm}
 F. Darabi$^2$\thanks{Email: f.darabi@azaruniv.edu (Corresponding author)   }
 \hspace{2mm}, and \hspace{2mm}
 H. Motavalli$^1$\thanks{Email: motavalli@tabrizu.ac.ir } \\
\centerline{$^1$\small {\em Department of Theoretical Physics
and Astrophysics, University of Tabriz, 51666-16471, Tabriz,
Iran.}}\\
{$^2$\small {\em Department of Physics, Azarbaijan Shahid Madani University , 53714-161, Tabriz, Iran. }}}

\maketitle
\begin{abstract}
We study the black hole thermodynamics and obtain the correction terms for temperature, entropy, and heat capacity of the Schwarzschild black hole, resulting from the commutation relations in the framework of {\it Modified Generalized Uncertainty Principle} suggested by {\it Doubly Special Relativity}.
\end{abstract}

\newpage
 
\section{Introduction}

The first idea that gravity may affect the quantum uncertainty principle dates back to Mead \cite{Mead}. This idea was based on the fact that in the strong gravity regime, Heisenberg uncertainty relation is no longer satisfactory. Since then, modified commutation relations between position and conjugate momentum commonly known as Generalized Uncertainty Principle (GUP) were introduced in the context of string theory and
black hole physics with the prediction of a minimum measurable length 
\cite{Gross,Kato,Kato1,Kato2,Kato3,Kato4,Kato5,Kato6,general GUP3,general1 GUP3,string1,string2,string3,string4,string5,string6,Kemp,Kemp1,Kemp2,Kemp3,Kemp4,Kemp5}.
A large amount of interest has recently been focused on
resolving the quantum corrections for Schwarzschild black hole thermodynamics, by GUP \cite{Amel}-\cite{Amel4}. Corrections to the thermodynamical properties of Reissner-Nordstr\"om black hole has been considered in \cite{Yoon}, and black holes in anti-de Sitter space in \cite{Setare,Setare1}. On the other hand, black hole thermodynamics with GUP has been studied in \cite{Jalal}-\cite{Jalal3} and the entropy of a charged black hole in f(R) gravity is studied in \cite{Adami}. In principle, one may think of the GUP as a model independent concept ideally perfect for the study of black hole thermodynamics \cite{Medved,Bolen}. It has also been discussed that the GUP may give rise to black hole remnants which may be considered as candidate for dark matter \cite{Adler,Chen,Nozari,Xiang,Nicol}. 

Recently, a modified GUP is proposed which is consistent with {\it Doubly Special Relativity} (DSR) theory, \cite{DSR1,DSR2,DSR3,DSR4}.
The fact that the leading order correction to the Heisenberg's uncertainty principle should be linear (and not quadratic) in the Planck length was proposed for the first time in \cite{Bambi}, and then some implications for the thermodynamics of black holes was briefly discussed in \cite{Xiang1} where it has been shown
that the first order correction to the standard black hole entropy should be linear in the black hole mass.

Motivated by the common interest on the thermodynamics of black hole, in the present paper, we aim to study in detail the Schwarzschild black hole thermodynamics and obtain the correction terms for temperature, entropy, and heat capacity of the black hole, resulting from the commutation relations in the context of modified GUP suggested by doubly special relativity, or the DSR-GUP \footnote{Throughout the paper we will consider $\hbar = G = c = k_B= 1$, where $G$ is the Gravitational constant, $c$ is the velocity oflight and $k_B$ is the Boltzmann constant.}.

\section{Generalized uncertainty principal}

The Generalized uncertainty Principal (GUP) is a generalization of
the well-known Heisenberg's uncertainty principal at Planck scale. It inevitably affects the dynamics of high energy systems like the early universe. In a one dimensional system, the simplest form of the GUP can be written as \cite{Kemp}-\cite{Kemp5}
\bea\label{e1}
 \delta x \delta p\geq\frac{\hbar}{2}\Bigg(1+\beta \ell_{Pl}^2(\delta p)^2 \Bigg),
 \eea
where $\ell_{Pl}\sim10^{-35}m$ is the Planck length and $\beta$ is a
constant of order unity. It is obvious that the new second
term in (\ref{e1}) is important only at Planck scale, i.e. at small
length and very high energy$(10^{16}TeV)$. Recently, the authors in
\cite{A.F.Ali1,A.F.Ali2} have discussed a GUP which is in
agreement with Doubly Special Relativity (DSR) theory \cite{DSR1}-\cite{DSR4}. By assuming that space coordinates commute with space coordinates and momenta commute with momenta as

\bea\label{e2}
  [x_i,x_j]=0,   &            &[p_i,p_j]=0,
 \eea
the following algebra is obtained for phase space

 \bea\label{e3}
  [x_i,p_j]=i \Bigg\{\delta_{ij}-\alpha_0 {\ell_{Pl}}\Bigg(p \delta_{ij}+
  \frac{p_i p_j}{p}\Bigg)+\alpha_0^2{\ell^2_{Pl}}(p^2 \delta_{i} +3p_i
  p_j)\Bigg\},
 \eea
 \bea\label{e4}
 \delta x \delta p\geq\frac{1}{2}\Bigg(1-2\alpha_0 {\ell_{Pl}}(\delta p)
 +4\alpha_0^2{\ell_{Pl}^2}(\delta p)^2 \Bigg),
 \eea
where $\alpha_0$ is assumed to be of order unity. To distinguish between the linear and second order terms in the Planck length, we
rewrite the GUP of equation (\ref{e4}) in a more general form 
\cite{Barun Majumder}

\bea\label{e5}
 \delta x \delta p\geq\frac{1}{2}\Bigg(1+\beta \ell_{Pl}(\delta p)+\alpha^2\ell_{Pl}^2(\delta p)^2 \Bigg).
 \eea
Here, $\beta$ as the coefficient of  $\ell_{Pl}$ and $\alpha$ as the coefficient of $\ell_{Pl}^2$ are constant. The coefficient $\beta$ indicates the effect of linear term in the Planck length, so setting $\beta=0$ and $\beta=-2\alpha_0,
\alpha^2=4\alpha_0^2$, gives back the ordinary GUP equation (\ref{e1}), and the DSR-GUP equation (\ref{e4}), respectively. By using equation (\ref{e5}) one may write the momentum uncertainty as \cite{Barun Majumder}

\bea\label{e6}
  \delta p&\geq&
  \frac{1}{2\delta x}f_{_{DSR-GUP}}(\delta x),
 \eea
where
 \bea\label{e6.1}
 f_{_{DSR-GUP}}(\delta x)&=&\Bigg[\frac{2(\delta x)^2}{\alpha^2\ell_{Pl}^2}\Bigg(1-\frac{\beta \ell_{Pl}}{2\delta x}\Bigg)
  \Bigg\{1\pm\sqrt{1-\frac{\alpha^2\ell_{Pl}^2}{(\delta x)^2(1-\frac{\beta \ell_{Pl}}{2\delta x})^2}}\Bigg\} \Bigg]
  .
   \eea
The requirement of reality for $f_{_{DSR-GUP}}(\delta x)$ implies the existence of a minimum measurable length 

 \bea\label{e7}
 \delta x\geq\delta
x_{min}=(2\alpha+\beta)\frac{\ell_{Pl}}{2},
 \eea
and a maximum measurable momentum, using (\ref{e6}), as
\bea\label{e8}
 \delta
p\leq\delta p_{max}=\frac{M_{Pl}}{\alpha}.
\eea

We see that $\delta p_{max}$ is independent of $\beta$, and for the ordinary GUP and DSR-GUP, $\delta x_{min}$ is equal to $\ell_{Pl}$.
In equation (\ref{e6.1}), we choose negative sign, because we want
to recover Bekenstein-Hawking temperature in the limit of large
$M_{BH}$ in the following section. For later use we write $f_{_{DSR-GUP}}(\delta
x)$ in series form in $\alpha$ and $\beta$ \cite{Barun Majumder}:

\bea\label{e6.5}
 f_{_{DSR-GUP}}(\delta x)
  &=&1+\frac{\beta \ell_{Pl}}{2\delta
  x}+\frac{(\alpha^2+\beta^2)\ell_{Pl}^2}{(2\delta x)^2}+\frac{(3\alpha^2\beta+\beta^3)\ell_{Pl}^3}{(2\delta
  x)^3}
  \nnb \\ &+&
  \frac{(2\alpha^4+\beta^4+6\alpha^2\beta^2)\ell_{Pl}^4}{(2\delta
  x)^4}+\frac{(10\alpha^4\beta+10\alpha^2\beta^3+\beta^5)\ell_{Pl}^5}{(2\delta
  x)^5}
  \nnb \\
  &+&
  \sum_{d=3}\left[\frac{f_{2d}(\alpha\beta)\ell_{Pl}^{2d}}{(2\delta
  x)^{2d}}+\frac{f_{2d+1}(\alpha\beta)\ell_{Pl}^{2d+1}}{(2\delta
  x)^{2d+1}}\right].
   \eea

\section{Black hole thermodynamics}

The entropy and the temperature of a stationary Schwarzschild black hole
are given respectively by \cite{Bekenstein1}-\cite{Hawking3}

\bea\label{e9}
S_{BH}=\frac{A_{BH}}{4\ell_{Pl}^2}, 
  \eea
  \bea\label{e9'}
T_{BH}=\frac{M_{Pl}^2}{8\pi M_{BH}},
  \eea
where $A_{BH}$ and $M_{BH}$ are the area and mass of the black hole, respectively.

Let us now assume that the apparent horizon of black hole radiates or absorbs a single photon with energy $dE$. We can identify this energy with the uncertainty in the momentum of the photon. If we apply the Heisenberg's uncertainty principle to this situation, the increasing or decreasing in temperature of the black hole is given by

\bea\label{e10}
 dT_{BH}=-\frac{M_{Pl}^2}{8\pi M^2_{BH}}dM=\frac{M_{Pl}^2}{8\pi M^2_{BH}}\delta p\approx\frac{M_{Pl}^2}{8\pi M^2_{BH}}\frac{1}{2\delta x},
  \eea
By considering (\ref{e6}), the increasing or decreasing in temperature of the black hole in the DSR-GUP framework then becomes

\bea\label{e11}
 dT_{_{DSR-GUP}}&\approx& \frac{M_{Pl}^2}{8\pi M^2_{BH}}\frac{1}{2\delta x}f_{_{DSR-GUP}}(\delta x)
 \nnb \\
 &\approx&f_{_{DSR-GUP}}(\delta x)dT_{BH}.
  \eea
We know that in the vicinity of black hole surface, the uncertainty in position
of an emitted photon is about twice the Schwarzschild radius. So, by using the definition of Schwarzschild radius and temperature of the black hole we obtain  

\bea\label{e12}
 \delta x\approx 2r_s=4 M_{BH}=\frac{\ell_{Pl}M_{Pl}}{2\pi T_{BH}},
  \eea
Note that, if we choose $ \delta x_{min}= 2(r_s)_{min}=4 M_{min}$ and use (\ref{e7}), we see that the minimum size and mass of the black hole are given by

\bea\label{e7.5}
 (r_s)_{min}=(2\alpha+\beta)\frac{\ell_{Pl}}{4}, &          &M_{min}=(2\alpha+\beta)\frac{M_{Pl}}{8}.
 \eea
By substituting $\delta x$ from (\ref{e12}) in equation (\ref{e11}) and using expanded form of $f_{_{DSR-GUP}}(\delta x)$ and then integrating, we find

\bea\label{e13}
 T_{_{DSR-GUP}}&=&T_{BH}+\frac{\pi\beta}{2M_{Pl}} T_{BH}^2+\frac{\pi^2(\alpha^2+\beta^2)}{3M_{Pl}^2}T_{BH}^3
 +\frac{\pi^3(3\alpha^2\beta+\beta^3)}{4M_{Pl}^3}T_{BH}^4
  \nnb \\ &+&
  \frac{\pi^4(2\alpha^4+\beta^4+6\alpha^2\beta^2)}{5M_{Pl}^4}T_{BH}^5
  +\frac{\pi^5(10\alpha^4\beta+10\alpha^2\beta^3+\beta^5)}{6M_{Pl}^5}T_{BH}^6
  \nnb \\
  &+&
  \sum_{d=3}\left[\frac{\pi^{2d}f_{2d}(\alpha\beta)}{(2d+1)M_{Pl}^{2d}}T_{BH}^{(2d+1)}
  +\frac{\pi^{(2d+1)}f_{2d+1}(\alpha\beta)}{(2d+2)M_{Pl}^{(2d+1)}}T_{BH}^{(2d+2)}\right]+\mbox{Const.}
   \eea
where the integration constant may be a function of
$\alpha$ and $\beta$. By using equation (\ref{e9'}), we can write
$T_{_{DSR-GUP}}$ as a function of black hole mass as

\bea\label{e13.5}
 T_{_{DSR-GUP}}&=&\frac{1}{8\pi}\frac{M_{Pl}^2}{ M_{BH}}+\frac{\beta}{2\times 8^2\pi}\frac{M_{Pl}^3}{ M_{BH}^2}
 +\frac{(\alpha^2+\beta^2)}{3\times 8^3\pi}\frac{M_{Pl}^4}{ M_{BH}^3}+\frac{(3\alpha^2\beta+\beta^3)}{4\times 8^4\pi}\frac{M_{Pl}^5}{ M_{BH}^4}
  \nnb \\ &+&
  \frac{(2\alpha^4+\beta^4+6\alpha^2\beta^2)}{5\times 8^5\pi}\frac{M_{Pl}^6}{ M_{BH}^5}
  +\frac{(10\alpha^4\beta+10\alpha^2\beta^3+\beta^5)}{6\times 8^6\pi}\frac{M_{Pl}^7}{ M_{BH}^6}
  \nnb \\
  &+&
  \sum_{d=3}\left[\frac{f_{2d}(\alpha\beta)}{(2d+1)8^{(2d+1)}\pi}\frac{M_{Pl}^{(2d+2)}}{ M_{BH}^{(2d+1)}}
  +\frac{f_{2d+1}(\alpha\beta)}{(2d+2)8^{(2d+2)}\pi}\frac{M_{Pl}^{(2d+3)}}{ M_{BH}^{(2d+2)}}\right]+\mbox{Const.}
   \eea
Here, we set $\mbox{Const}=0$ so as to make the result be in agreement with the ordinary one $(\ref{e9'})$, for large mass. The temperature as a function of mass of black hole is shown in figure 1.left. In the GUP framework, the
temperature is greater than ordinary one but in the context of DSR-GUP,
it is lesser due to $\beta<0$. This is a consequence of the linear term in Planck length in the modified uncertainty relation which causes more
stability of black hole, especially with little mass (see figure
2.rigth).

Now, we calculate the entropy in the context of DSR-GUP. As we
mentioned above, emission or absorbtion of a photon changes the area of
black hole as well as its entropy. By using equation (\ref{e9}) and $T
dS= dE$, we have

\bea\label{e14}
 dA=\frac{4}{T} dE.
   \eea
If we apply Heisenberg uncertainty relation, we obtain
\bea\label{e15}
 dA_{BH}=\frac{4}{T} \frac{1}{2\delta x}.
   \eea
In the DSR-GUP framework, this becomes as follows

\bea\label{e16}
 dA_{_{DSR-GUP}}&=&\frac{4}{T} \frac{1}{2\delta x}f_{_{DSR-GUP}}(\delta x)
\nnb \\
 &=&dA_{BH}f_{_{DSR-GUP}}(\delta x).
   \eea
 Now, using $\delta x\approx2r_s=2(\frac{A_{BH}}{4\pi})^{\frac{1}{2}}$
 and inserting it in equation (\ref{e16}) and integrating, results in

\begin{figure}
\vskip 1.5 cm
    \includegraphics{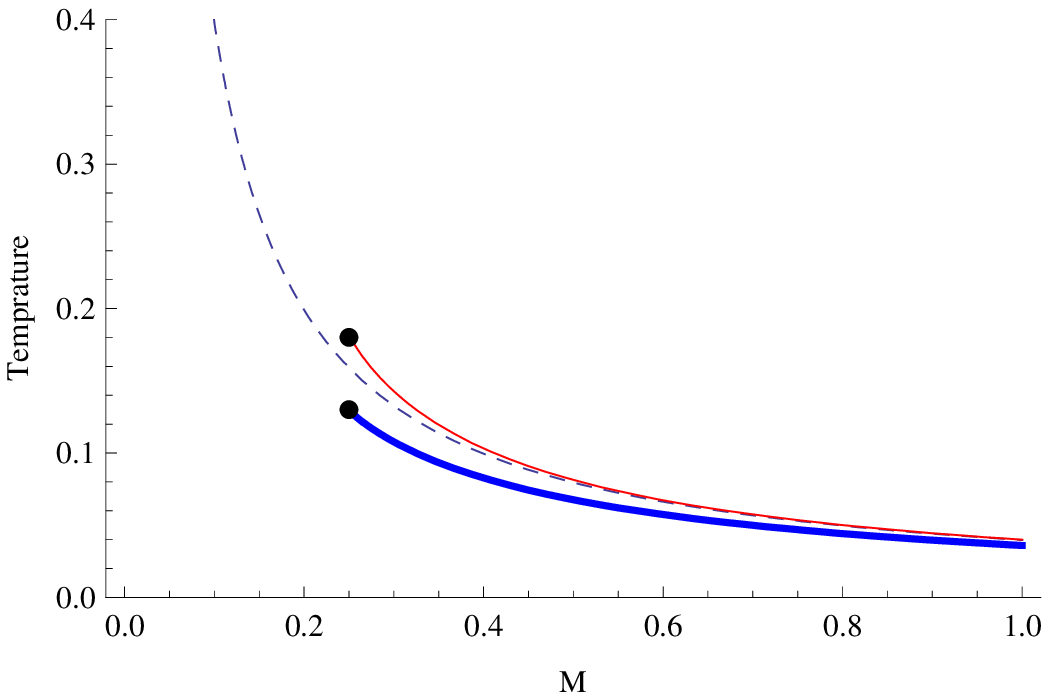}
     \includegraphics{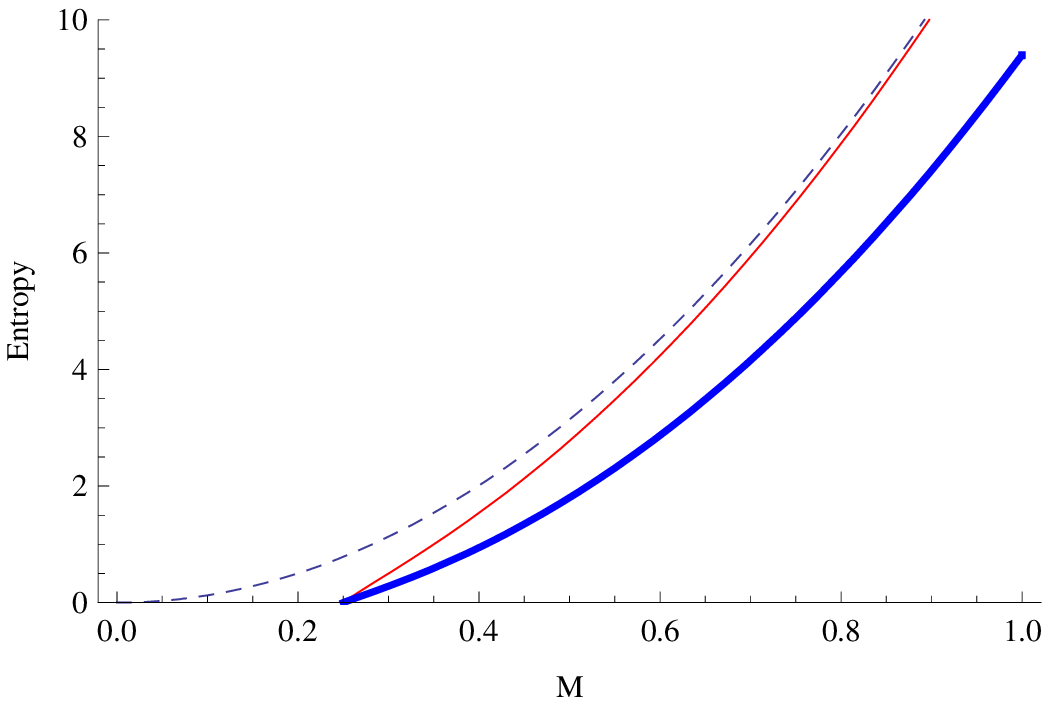}
\vskip 2.5 cm \caption{ Left and right figures show the
dependence of temperature and the entropy of a black hole on
mass, respectively. Mass and temperature are expressed in units of the Planck mass and Planck energy, respectively. The dashed, red-thin and blue-thick curves are drawn in the ordinary, GUP and DSR-GUP frameworks, respectively.}
\end{figure}

\bea\label{e17}
 A_{_{DSR-GUP}}
  &=&A_{BH}+(\pi)^{\frac{1}{2}}\beta \ell_{Pl}(A_{BH})^{\frac{1}{2}}+\pi\frac{(\alpha^2+\beta^2)\ell_{Pl}^2}{4}
  lnA_{BH}-
  (\pi)^{\frac{3}{2}}\frac{(3\alpha^2\beta+\beta^3)\ell_{Pl}^3}{4}(A_{BH})^{-\frac{1}{2}}
  \nnb \\ &-&
  \pi^2\frac{(2\alpha^4+\beta^4+6\alpha^2\beta^2)\ell_{Pl}^4}{16}(A_{BH})^{-1}-
  (\pi)^{\frac{5}{2}}\frac{(10\alpha^4\beta+10\alpha^2\beta^3+\beta^5)\ell_{Pl}^5}{48}(A_{BH})^{-\frac{3}{2}}
  \nnb \\
  &-&
  \sum_{d=3}\left[\frac{\pi^df_{2d}(\alpha\beta)\ell_{Pl}^{2d}}{(d-1)2^{2d}}(A_{BH})^{-d+1}+
  2\frac{\pi^{\frac{2d+1}{2}}f_{2d+1}(\alpha\beta)\ell_{Pl}^{2d+1}}{(2d-1)2^{2d+1}}(A_{BH})^{-\frac{2d-1}{2}}\right]+\mbox{Const},
   \eea
where the integration constant may be a function of
$\alpha$ and $\beta$. Using Bekenstein-Hawking relation for the
entropy, namely Eq.(\ref{e9}), we can calculate the DSR-GUP entropy
as a function of the mass of black hole 

\bea\label{e18}
 S_{_{DSR-GUP}}
  &=&\frac{4\pi M_{BH}^2}{M_{Pl}^2}+\pi \beta \frac{M_{BH}}{M_{Pl}}+\frac{\pi}{8}(\alpha^2+\beta^2)
  ln\frac{ M_{BH}}{M_{Pl}}-
  \frac{\pi}{64}(3\alpha^2\beta+\beta^3)\frac{M_{Pl}}{M_{BH}}
  \nnb \\ &-&
  \frac{\pi}{1024}(2\alpha^4+\beta^4+6\alpha^2\beta^2)\frac{M_{Pl}^2}{M_{BH}^2}-
  \frac{\pi}{12288}(10\alpha^4\beta+10\alpha^2\beta^3+\beta^5)\frac{M_{Pl}^3}{M_{BH}^3}
  \nnb \\
  &-&
  \sum_{d=3}\left[\frac{\pi}{2^{6d-2}}\frac{f_{2d}(\alpha\beta)}{(d-1)}(\frac{M_{Pl}}{M_{BH}})^{2d-2}+
  \frac{\pi}{2^{6d}}\frac{f_{2d+1}(\alpha\beta)}{(2d-1)}(\frac{M_{Pl}}{M})^{2d-1}\right]+\mbox{Const.}
   \eea
Here, we set the constant of integration so that when the mass of black
hole goes to its minimum $M_{min}$, the entropy reaches to
zero. We see a new correction term $\sim M_{BH}$, added to the
entropy, which is a consequence of the linear term in Planck length
in the modified uncertainty principal. As is seen in Fig.1 (right), both
of the (blue-thick) curve corresponding to the DSR-GUP and the (red-thin) curve corresponding to the GUP deviate drastically from the (dashed) curve corresponding to the ordinary framework, for smaller masses. However, the GUP curve approaches the ordinary curve for larger masses, whereas the DSR-GUP curve deviates drastically from the ordinary curve at larger masses. Also, in Fig.1 (left), both of the (blue-thick) curve corresponding to the DSR-GUP and the (red-thin) curve corresponding to the GUP approach the ordinary curve, for larger masses, but deviate from it for smaller masses. 

Now, we may calculate the time dependence of mass of the black hole. To do
this, we assume that the energy loss is dominated by photon and
then use the Stefan-Boltzmann law. For the ordinary case this leads
to

\begin{figure}
\vskip 1.5 cm
    \includegraphics{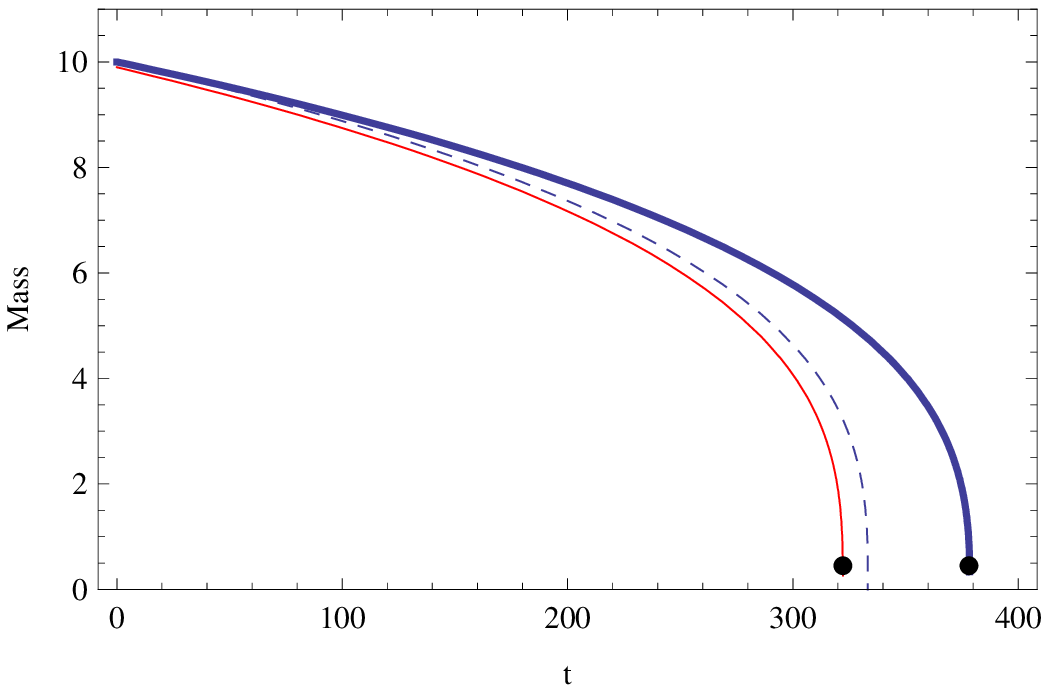}
    \includegraphics{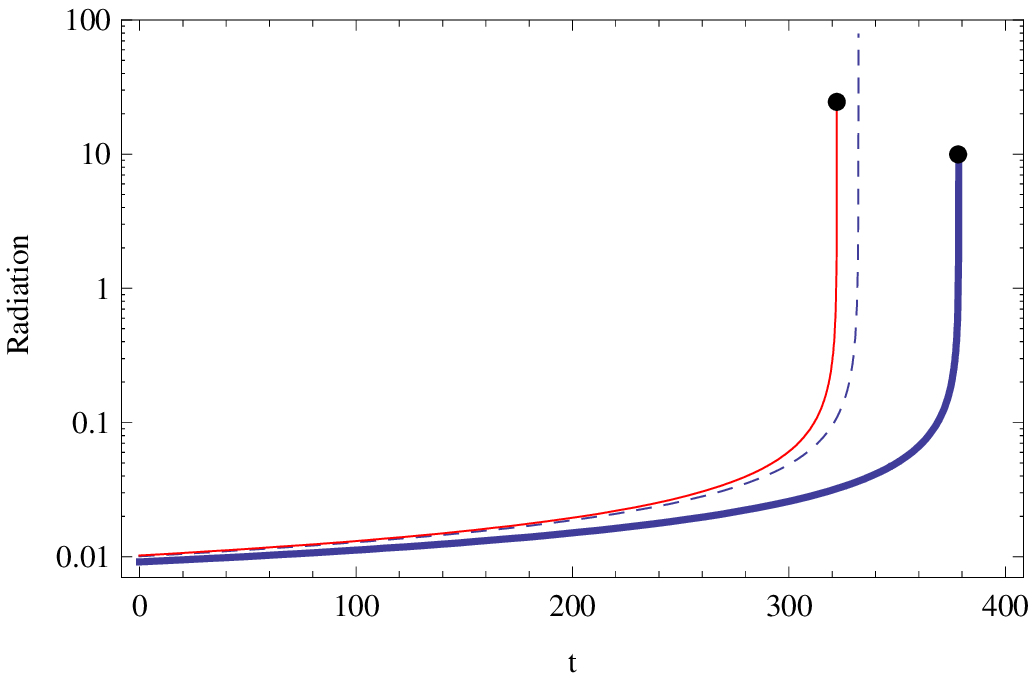}
\vskip 2.5cm \caption{Left and right figure are respectively the
time dependence of the mass and the radiation of a black hole with
$M_{BH}=10M_{Pl}$. Time is in the unit of characteristic time,
mass is in the unit of Planck mass and radiation is in the unit of Planck energy per characteristic time. The dashed, red-thin and blue-thick curves are drawn in the ordinary, GUP and  DSR-GUP frameworks, respectively.}
\end{figure}

 \bea\label{e19}
\frac{d}{dt}(\frac{M_{BH}}{M_{Pl}})=-\frac{1}{60\times16^2\pi
T_{Pl}}(\frac{M_{Pl}}{M_{BH}})^2,  & &T_{Pl}=\frac{\hbar}{M_{Pl}}
 \eea

\bea\label{e20}
\frac{M_{BH}}{M_{Pl}}=\Bigg[\Bigg(\frac{M_i}{M_{Pl}}\Bigg)^3-\frac{3t}{t_{ch}}\Bigg]^\frac{1}{3},
&          &  t_{ch}=60\times16^2\pi T_{Pl}
 \eea

\bea\label{e21}
\frac{d}{dt}(\frac{M_{BH}}{M_{Pl}})=-\frac{1}{t_{ch}}\Bigg[\Bigg(\frac{M_i}{M_{Pl}}\Bigg)^3-\frac{3t}{t_{ch}}\Bigg]^{-\frac{2}{3}}.
   \eea
where $M_i$ is the initial mass of black hole and $t_{ch}$ is the
characteristic time which is about $4.8\times10^4$ times the
Planck time. The black hole evaporates to zero mass within the time duration
$t=\frac{t_{ch}}{3}(\frac{M_i}{M_{Pl}})^3$ but the radiation rate
is infinite at the end of the process.

In the GUP and DSR-GUP frameworks, by numerical investigation
of Stefan-Boltzmann law, we see that the black hole evaporates
to $M_{BH}=\frac{M_{Pl}}{4}$ as a remnant (Figure
2.left). In this case, the energy radiation rate is finite at the
end point (Figure 2.right). Also the evaporation time is different
from ordinary one. In the context of GUP, the black hole reaches to
the end point sooner than ordinary one with ignorable time difference, whereas in the DSR-GUP framework the black hole evaporation time is longer than ordinary one and the time difference is considerable.

According to the figures 1 and 2, the temperature and radiation
of the black hole is negligible for large-mass black holes. The
black hole absorbs much more from the microwave background
radiation than it radiates itself as Hawking radiation. Therefore, one can definitely say that the Hawking radiation plays no important role in the case of large-sized black holes. The only type of black hole where one can hope to observe Hawking radiation is the so-called mini black hole, which might have been existed in the primordial stage of Universe. These primordial mini black holes radiate and leave remnants. These remnants can not radiate further and have only gravitational interactions rendering them as a candidate of dark matter.

\begin{figure}
\vskip 1.5 cm
    \includegraphics{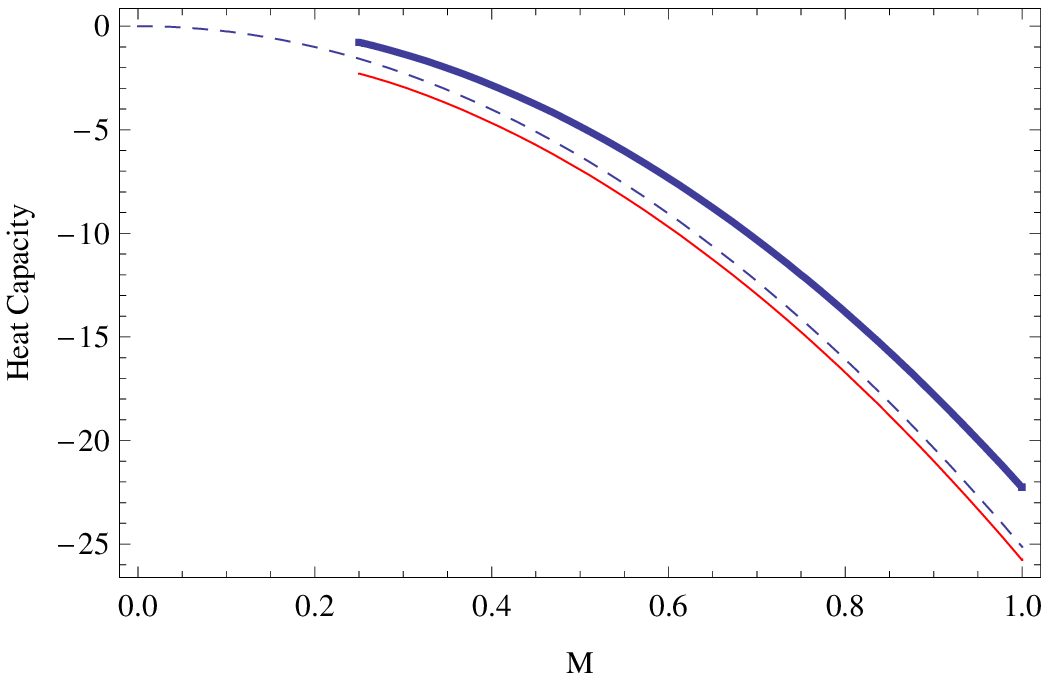}
    \includegraphics{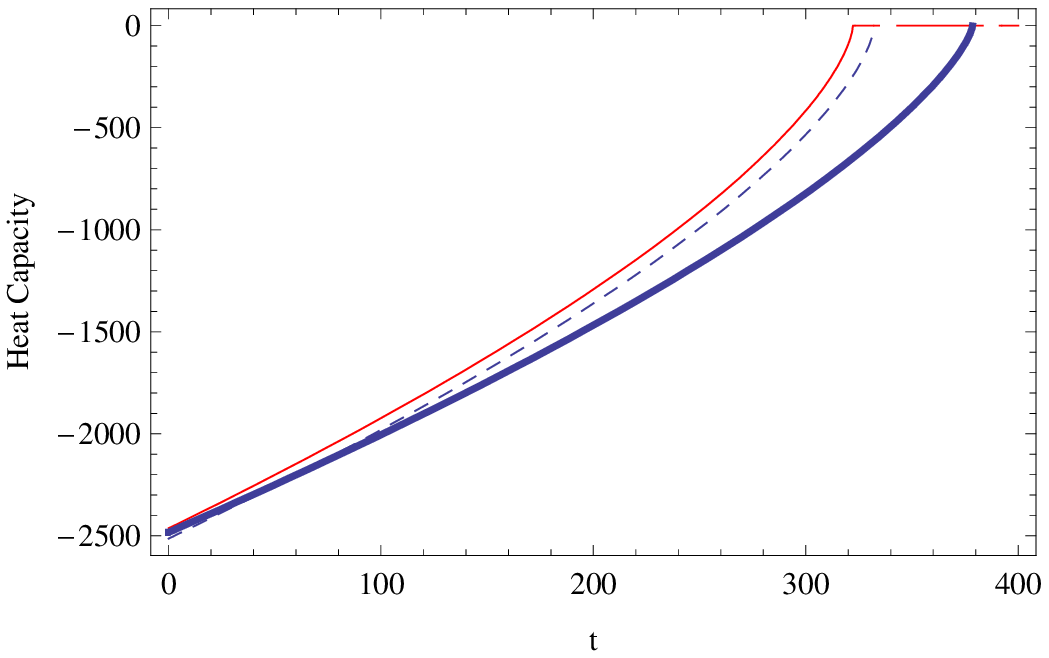}
\vskip 2.5cm \caption{Left figure is the heat capacity of a black
hole versus mass and the right is the time dependence of the heat
capacity of a black hole with $M_{BH}=10M_{Pl}$. Time is in the
unit of characteristic time and mass is in the unit of Planck mass. The dashed, red-thin and blue-thick curves are drawn in the ordinary, GUP and  DSR-GUP frameworks, respectively.}
\end{figure}

Finally, we check the influence of GUP in the heat
capacity of black hole. In the ordinary case, using $C=dE/dT$,
$TdS=dE$ and equations (\ref{e9'}), (\ref{e20}) we have

\bea\label{e22}
C_{BH}=-\frac{8\pi M_{BH}^2}{M_{Pl}^2}
=-8\pi\Bigg[\Bigg(\frac{M_i}{M_{Pl}}\Bigg)^3-\frac{3t}{t_{ch}}\Bigg]^\frac{2}{3}.
 \eea
To derive the heat capacity in the context of the DSR-GUP, we use

\bea\label{e23} C_{BH} dT=dE=\frac{1}{2\delta x}\:,
 \eea
so that we can write
\bea\label{e23'}
C_{_{DSR-GUP}}dT=dE=f_{_{DSR-GUP}}(\delta x)\frac{1}{2\delta x}.
 \eea
By comparing Eqs.(\ref{e23}), (\ref{e23'}), and then using equations (\ref{e12}), (\ref{e22}) we have

\bea\label{e24}
 C_{_{DSR-GUP}}&=&f_{_{DSR-GUP}}(M_{BH})C_{BH}
  \nnb \\&=&
  -8\pi\Bigg(\frac{M_{BH}^2}{M_{Pl}^2}+\frac{\beta
M_{BH}}{8M_{Pl}}+\frac{(\alpha^2+\beta^2)}{(8)^2}+\frac{(3\alpha^2\beta+\beta^3)M_{Pl}}{(8)^3M_{BH}}
  \nnb \\ &+&
  \frac{(2\alpha^4+\beta^4+6\alpha^2\beta^2)\ell_{Pl}^2}{64(8M_{BH})^2}+\frac{(10\alpha^4\beta+10\alpha^2\beta^3+\beta^5)\ell_{Pl}^3}{64(8M_{BH})^3}
  \nnb \\
  &+&
  \sum_{d=3}\left[\frac{f_{2d}(\alpha\beta)M_{Pl}^{2d-2}}{64(8M_{BH})^{2d-2}}+\frac{f_{2d+1}(\alpha\beta)M_{Pl}^{2d-1}}{64(8M_{BH})^{2d-1}}\right]\Bigg).
 \eea
Here, we see a correction term $\sim M_{BH}$ which is a consequence of the linear term in Planck length in the DSR-GUP. Figure 3, shows the heat capacity of a black hole versus mass, and also the time dependence of heat
capacity of a black hole with mass $M_{BH}=10M_{Pl}$. As is seen in Fig.3 (left), both of the (blue-thick) curve corresponding to the DSR-GUP and the (red-thin) curve corresponding to the GUP deviate from the (dashed) curve corresponding to the ordinary framework, for smaller masses. However, the GUP curve approaches the ordinary curve for larger masses, whereas the DSR-GUP curve deviates drastically from the ordinary curve at larger masses.

\newpage

\section{Conclusion}

We have studied the black hole thermodynamics and obtained the correction terms for temperature, entropy, and heat capacity of black hole in the framework of modified generalized uncertainty principle suggested by {\it Doubly Special Relativity} theory, namely (DSR-GUP). 
It is shown that while in the GUP framework the temperature is greater than ordinary one, in the DSR-GUP framework the temperature is less than the ordinary one, especially for smaller black holes. This is a consequence of the linear term in Planck length appearing in the modified uncertainty principle.
We have also obtained a new correction term $\sim M_{BH}$ added to the
entropy, in the DSR-GUP framework which is again a consequence of the linear term in Planck length in the modified uncertainty principal. The study of time dependence of the black hole's mass shows that in the GUP and DSR-GUP frameworks the black hole evaporates to $M_{BH}=\frac{M_{Pl}}{4}$ as a remnant. In the context of GUP, black hole reaches to the end point sooner than ordinary one, but in the DSR-GUP framework the black hole evaporation time is longer than ordinary one. Finally, we have studied the influence of DSR-GUP on the heat capacity of black hole and found a correction term $\sim M_{BH}$.

\end{document}